\documentclass[dvipsnames]{arxivcls}
\usepackage{arxiv}
 
\arxivSubmission
\usepackage[T1]{fontenc}
\usepackage{dfadobe}
\usepackage{cite}
\BibtexOrBiblatex
\electronicVersion
\PrintedOrElectronic
\ifpdf \usepackage[pdftex]{graphicx} \pdfcompresslevel=9
\else \usepackage[dvips]{graphicx} \fi

\usepackage{algorithm}
\usepackage{algpseudocode}
\usepackage{pifont}
\usepackage{egweblnk}
\usepackage{amsmath}
\usepackage{todonotes}
\usepackage{subcaption}
\usepackage{pgfplotstable}
\usepackage{overpic}
\usepackage{multirow}
\usepackage{svg}
\usepackage{tikz}
\usepackage[all]{nowidow}
\usetikzlibrary{patterns}
\usetikzlibrary{calc, tikzmark}
\usepgfplotslibrary{fillbetween}
\usepackage{color}

\title[Progressive Mesh Unfolding]%
      {Unfolding via Progressive Mesh Approximation}

\author[L. Zawallich \& R. Pajarola]
{
  \parbox{.4\textwidth}{\centering Lars Zawallich\orcid{0000-0001-5730-4361}\\
                                 University of Zurich, Department of Informatics, Switzerland\\
                                 \url{LarsZawallich@gmail.com}
          }
  \hfill
  \parbox{.4\textwidth}{\centering Renato Pajarola\orcid{0000-0002-6724-526X}\\
                                 University of Zurich, Department of Informatics, Switzerland\\
                                 \url{pajarola@ifi.uzh.ch}
          }
}

\begin{document}
  \maketitle

\begin{abstract}
  When folding a 3D object from a 2D material like paper, typically only an approximation of the original surface geometry is needed.
  Such an approximation can effectively be created by a (progressive) mesh simplification approach, e.g. using an edge collapse technique.
  Moreover, when searching for an unfolding of the object, this approximation is assumed to be fixed.
  In this work, we take a different route and allow the approximation to change while searching for an unfolding.
  This way, we increase the chances to overcome possible ununfoldability issues.
  To join the two concepts of mesh approximation and unfolding, our work combines the edge collapsing mesh simplification technique with a \emph{Tabu Unfolder}, a robust mesh unfolding approach.
  We empirically show that this strategy performs faster and that it is more reliable than prior state of the art methods.


\end{abstract}

\section{Introduction}
\label{sec:Introduction}

In recent years, folding objects -- e.g. do-it-yourself decorations -- from flat materials like paper has gained increasing commercial attention.
Moreover, (un)folding is used in e.g. architecture~\cite{ArchitecturalFolding}, decorations~\cite{PCBend}, or robotics~\cite{OrigamiRobots}.
The objects to fold in these applications typically are an approximation of a high resolution model.
A typical manufacturing pipeline for such objects can be described by the following steps~\cite{SheetMetalBending}:

\begin{enumerate}
  \item Approximate a 3D object with a discrete surface.
  \item Unfold the approximation.
  \item Cut out the unfolding.
  \item Refold it (possibly automated).
\end{enumerate}

The unfolding needs to be overlap-free to cut it from a 2D material.
Moreover, it is of great efficiency advantage, if the unfolding is also single-patched~\cite[Section 22.1.1]{GeometricFoldingAlgorithms}.

Commonly, in the above mentioned pipeline, each step is treated separately.
Especially the unfolding (Step~2) is commonly done independently of the other steps.

One of the most intuitive techniques to unfold a polyhedron is edge unfolding (see Section~\ref{sec:Background}).
Even though it is proven that some non-convex polyhedra can not be unfolded using edge unfolding~\cite[Table 22.1]{GeometricFoldingAlgorithms}, in practice this problem rarely occurs.
However, there is yet no known general way to check if a polyhedron is edge-unfoldable or not.
Therefore, it is impossible to determine if an approximation of a 3D shape is in fact unfoldable apriori.
Even though the issue rarely occurs, it may still happen that the result of the first pipeline step is not edge-unfoldable.
Approaches treating each pipeline step individually, either ignore this issue, or try to solve it solely within the unfolding step.
To prevent confusion, we will call the result of Step~1 of the above mentioned pipeline the initial \emph{representation} of an object and any simplified approximation of this an \emph{approximation}.

We want to take a different approach than previous works did.
Instead of accepting a fixed representation, we allow variable approximations of the representation as results.
One commonly used simplification technique to create the initial representation is edge collapsing.
In this paper, we use edge collapses to simplify the representation even more, to a low resolution approximation.
This low resolution approximation has an expectable reduced unfold complexity while still having almost the same geometric shape as the desired result would have.
After unfolding the low face count approximation, we undo the mesh simplification operations until we reach the desired high resolution face count.
Since the geometry of the model is supposed to not change very much during this reverse process, little new unfolding overlaps are expected to occur.
Nevertheless, if during the reverse process an intermediate result fails to unfold, we return the last unfoldable approximation as a result.
We argue that since the initial representation is already an approximation, a slightly different approximation can be acceptable within most applications.

\section{Related Work}
\label{sec:RelatedWork}

Each unfolding of a polyhedron can also be seen as a distortionless parameterization.
Hormann et. al~\cite{ParameterizationCourse07} gave an overview of the field.
Current parameterization approaches aim to improve usability~\cite{BFF}, or interactivity~\cite{InteractiveParam}.

With respect to fabrication, developable surfaces (e.g. \cite{DevelopableI, DevelopableII, DevelopableIII}) are one way to create approximative paper crafts.
Instead of folding along given edges of the polyhedron, the goal of these methods is to create large bendable patches with zero Gaussian curvature, which approximate a given input shape.
In contrast to developable surfaces, our approach uses creases and folds, instead of bends and also creates a single-patched unfolding.

If not stated otherwise, any approach named below uses edge unfolding (see Section~\ref{sec:BackgroundUnfolding}) as a core technique.

Within the field of unfolding, several heuristic approaches have been explored to either create cut-trees~\cite{CreatingOptimizedCutOutSheets}, or unfold-trees~\cite{OptimalStrategiesForCreatingPaperModels}.
The core idea in both areas is to apply a weight to a main/dual-edge of a mesh and to then create a minimum spanning tree via well known methods, like Prim's or Kruskal's algorithm.
Both strategies then use segmentation to resolve remaining overlaps.

Xi et al.~\cite{InflationUnfolding} developed an approach to make polyhedra easier to unfold.
The core idea is to inflate and then to segment the mesh, to reduce local concavities which pose some of the key problems for unfolding.
This approach can be considered approximative.
However, in contrast to their method, we aim to create a single-patched unfolding.

One prominent approach aiming to create a single-patched unfolding was presented by Takahashi et al.~\cite{OptimizedTopologicalSurgery}.
They merge triangles using a genetic algorithm until only a single patch is left, or no overlap-free merge is possible anymore.
In the latter case, they return a segmented result, which our method always avoids.

Korpitsch et al.~\cite{SimulatedAnnealingUnfolding} applied simulated annealing to the problem of unfolding, while also considering glue-tags on top.
However, their approach is unable to handle not-unfoldable polyhedra and scales poorly~\cite[Section~5]{TabuUnfolding}.

Another approach aiming to create a single-patched unfolding is the Tabu Unfolding~\cite{TabuUnfolding} method (see Section~\ref{sec:BackgroundTabuUnfolding}).
The algorithm performs faster and more reliably than other comparable methods, but it is still unable to handle not-unfoldable input.

Our presented approach tries to overcome the not-unfoldability problem, by simplifying the input first, then unfolding the low resolution version and to uncollapse afterwards, while keeping the unfolding overlap-free.
In contrast to all other previous work, our approach is able to work with not-unfoldable input, while keeping the unfolding single-patched.

\section{Background and Definitions}
\label{sec:Background}

In this section we review the most important definitions and concepts related to unfolding.
Many definitions and explanations can also be found in the book \emph{Geometric Folding Algorithms: Linkages, Origami, Polyhedra}~\cite{GeometricFoldingAlgorithms}.

\subsection{Unfolding}
\label{sec:BackgroundUnfolding}

Cutting a polyhedron open and unfolding it such that all faces end up in a single common plane is called \emph{unfolding}.
Unfolding is a distortionless operation, i.e. no faces change their shape or size during unfolding.
Oftentimes, the term \emph{unfoldable} is used synonymous to ``unfoldable without self-overlap''.

\begin{figure}[ht]
  \begin{subfigure}{0.46\linewidth}
    \centering
    \includegraphics[width=\linewidth]{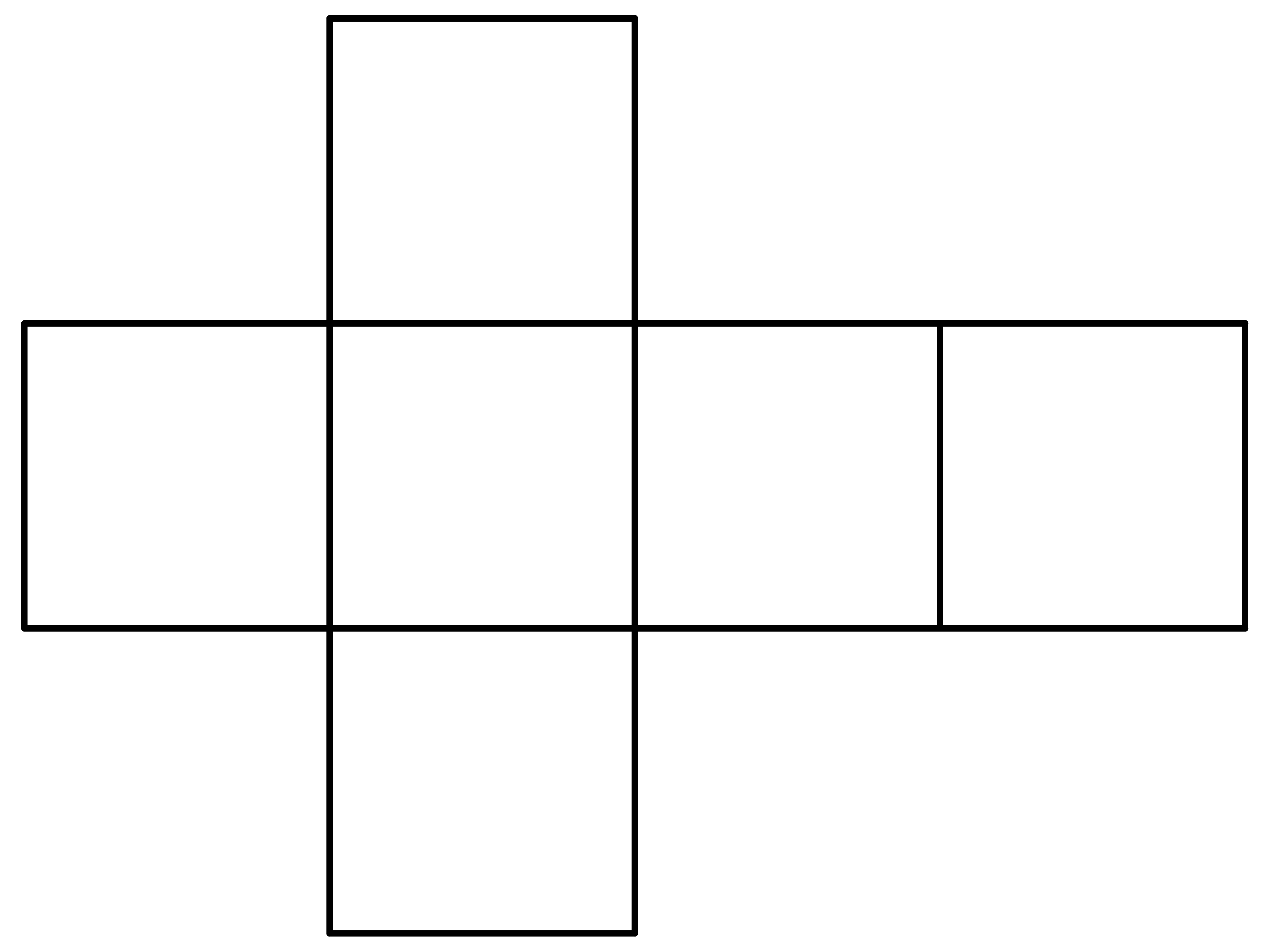}
    \caption{An unfolding of a cube, generated via edge unfolding.}
    \label{fig:EdgeUnfoldedCube}
  \end{subfigure}
  \hfill
  \begin{subfigure}{0.44\linewidth}
    \centering
    \includegraphics[width=\linewidth]{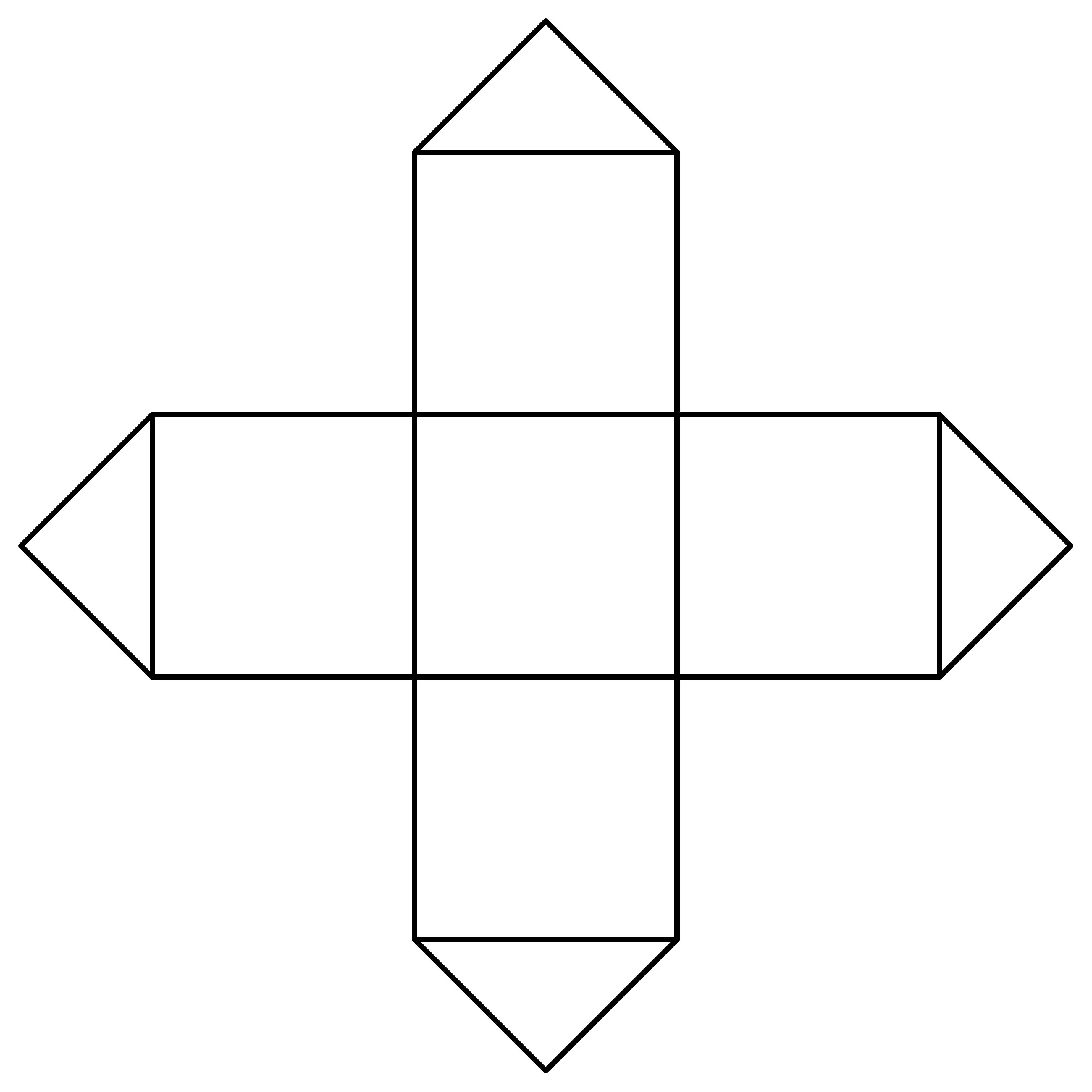}
    \caption{An unfolding of a cube, generated via general unfolding.}
    \label{fig:GeneralUnfoldedCube}
  \end{subfigure}
  \caption{Two different unfoldings of a cube.}
  \label{fig:EdgeVsGeneralUnfolding}
\end{figure}

Cuts may either happen only along the edges of the polyhedron or arbitrarily.
In the first case, the method is called \emph{edge unfolding}, in the second it is called \emph{general unfolding}.
Two exemplary cases of both methods for unfolding a cube are given in Figure~\ref{fig:EdgeVsGeneralUnfolding}.
An overlap-free single-patched unfolding created via edge unfolding is called a \emph{net}.
If a polyhedron does not have a net, it is called \emph{not-unfoldable}, or \emph{ununfoldable}.
In Table~\ref{tab:StateOfTheUnfolding}, an overview of the current state of research about unfoldability is given.
Within our work, we only use edge unfolding.

\begin{table}[ht]
  \begin{center}
    \begin{tabular}{l|cc}
       & \textbf{Edge unfolding} & \textbf{General unfolding} \\
      \hline
      \textbf{Convex} & open & always possible \\
      \textbf{Non-convex} & not always possible & open
    \end{tabular}
    \caption[Status of main questions concerning non-overlapping unfoldings]{Status of main questions concerning non-overlapping unfoldings.\protect\cite[Table 22.1]{GeometricFoldingAlgorithms}}
    \label{tab:StateOfTheUnfolding}
  \end{center}
\end{table}

\begin{figure}[ht]
  \centering
  \begin{subfigure}{.45\linewidth}
    \includegraphics[width=\textwidth]{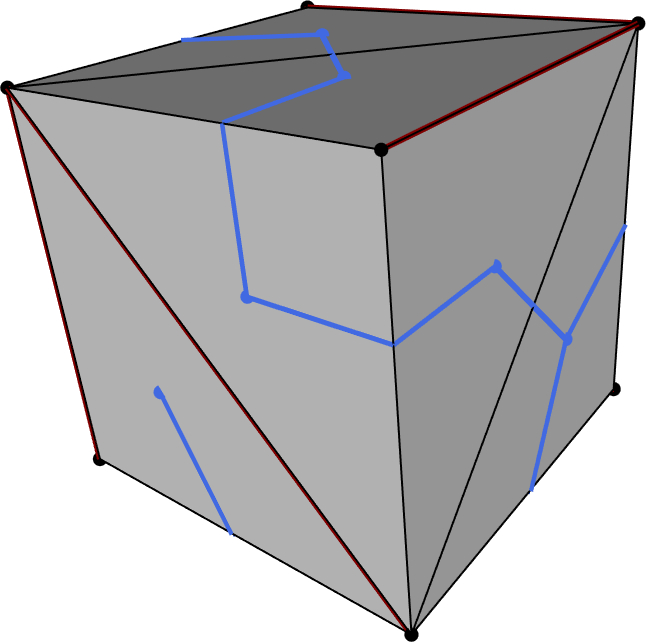}
    \caption{A folded cube.}
    \label{fig:FoldedCube}
  \end{subfigure}
  \hfill
  \begin{subfigure}{.45\linewidth}
    \includegraphics[width=\textwidth]{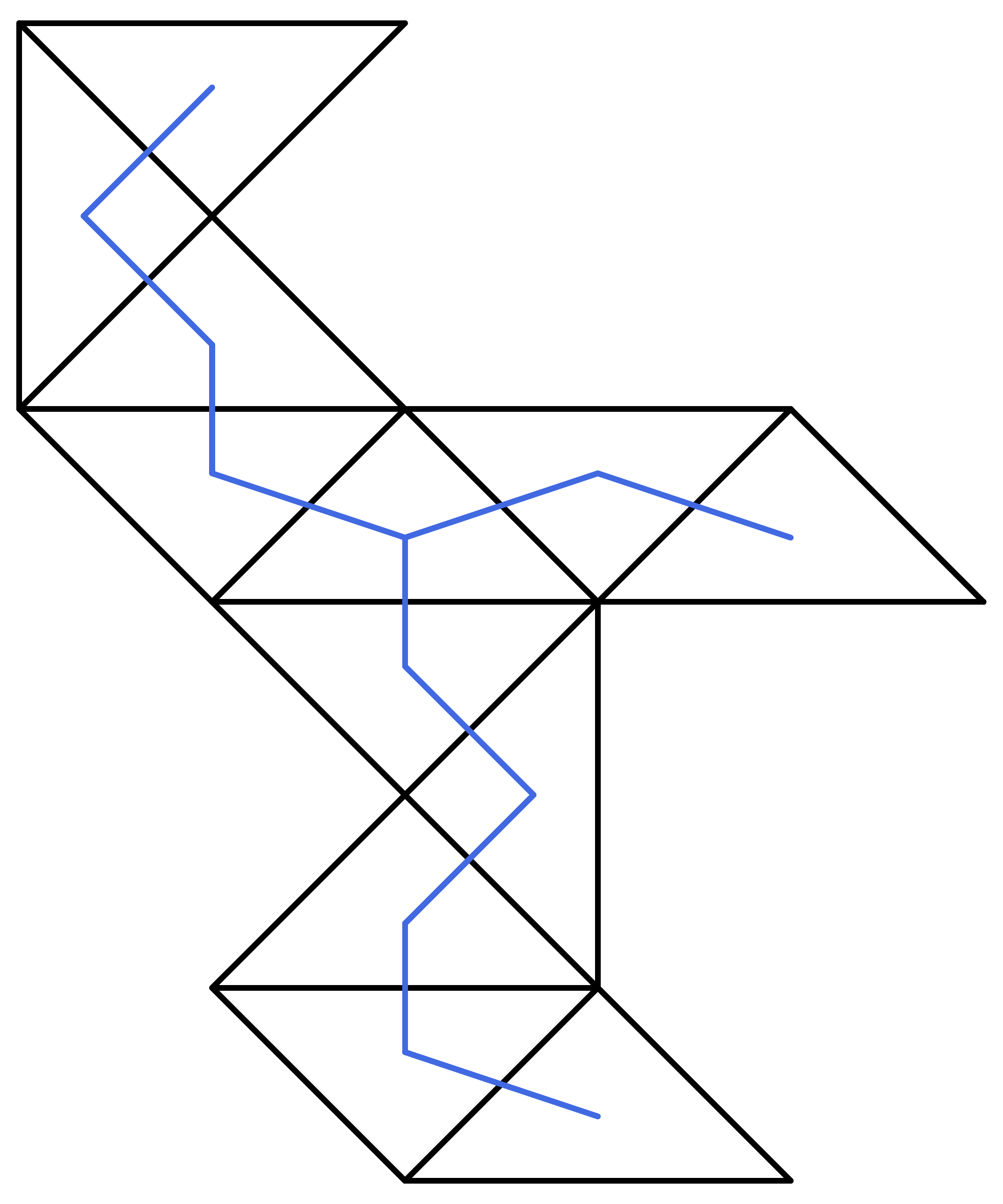}
    \caption{The corresponding unfolding.}
    \label{fig:UnfoldedCube}
  \end{subfigure}
  \caption{A folded triangulated cube and the corresponding unfolding. The unfold-tree is indicated in blue. The cut-tree over the edges of the mesh is visualized in dark red. In the unfolding, the cut-tree is the boundary and is not colored.}
  \label{fig:CutTreeAndUnfoldTree}
\end{figure}

There are two ways to define an unfolding, either via the cuts done or via the remaining connections of the faces.
In case the unfolding is defined via the cuts, these cuts form a spanning-tree over the graph of the polyhedron, if the mesh is of genus zero.
This spanning-tree is also called a \emph{cut-tree}.
If the mesh has a higher genus, the ``cut-tree'' needs to have equally many loops as the genus of the polyhedron.
In such a case, we call the structure a \emph{cut-graph}.
Unfortunately, the literature is not consistent and sometimes uses the term cut-tree also for non-genus zero objects.
An example unfolding including the cut-tree highlighted on the mesh is shown in Figure~\ref{fig:CutTreeAndUnfoldTree}.

In case the unfolding is defined via the remaining connections of the faces, these connections always form a spanning-tree over the dual-graph of the polyhedron.
We call such a spanning-tree an \emph{unfold-tree}.
In theory, areas composed of coplanar faces do not require cuts through them, which would allow for a loop in the unfold-tree.
Within our work, we still do the cut to increase flexibility while solving overlaps.
Each node only needs to hold a local transformation rotating the face it represents into the plane of its parent along their shared edge.
Unfolding the polyhedron is then done via traversing the tree once, while accumulating the local transformations downward.
In Figure~\ref{fig:CutTreeAndUnfoldTree} the unfold-tree is shown on the original as well as the unfolded mesh.

\subsection{Tabu Unfolding}
\label{sec:BackgroundTabuUnfolding}

To resolve overlaps, we use the \emph{Tabu Unfolding}~\cite{TabuUnfolding} approach, which we want to recap here.
The algorithm uses tabu search~\cite{TabuSearch} to iteratively reconfigure the unfold-tree, until no overlaps are left.
Note that any possibly self-overlapping unfolding can be used as a starting point.
The initial unfolding is created using the \emph{Steepest Edge Unfolder}~\cite{NetsOfPolyhedra}.
In each iteration, the method attaches a face in the unfolding to a new parent.
This action is called a \emph{move}.

In each iteration, the algorithm picks a random overlapping face.
If this face can be moved such that the overall number of overlaps is reduced, the move is performed.
Else, the algorithm recursively climbs the tree and does the same test, until a move is found which reduces the overall number of overlaps, or until the root node is reached.
In the latter case the best move found on the way to root is performed.

The relative best moves described above are important.
They allow to overcome plateaus or local minima.
To prevent the algorithm to fall back into a local minimum, it remembers the last $m$ moves and prevents these from being undone.
If all possible moves are on the tabu list, the tabu list is cleared, to re-enable moves.
As in the original work, we use $m = val \cdot \log_{val}(|F|)$, with $val$ being the average valence of the dual graph of the mesh and $|F|$ being the number of faces in the mesh.

While a tree-structure provides benefits like the easy unfolding by traversing the tree described above, it also poses some problems when resolving overlaps~\cite[Section~4.5]{TabuUnfolding}.
These problems include inefficiency, but also deadlocks, where the algorithm can not perform any more moves.
To prevent and resolve such issues Tabu Unfolding reroots the tree in each iteration.
Please note that the resulting unfolding layout does not change by rerooting the unfold-tree.
By rerooting, the algorithm can mimic the behavior of a flexible unfold-pattern, while maintaining the benefits of a tree-structure.
Within this work, we will refer to this method as \emph{TU}.

\section{Methods}
\label{sec:Methods}

\begin{figure*}[ht]
  \input{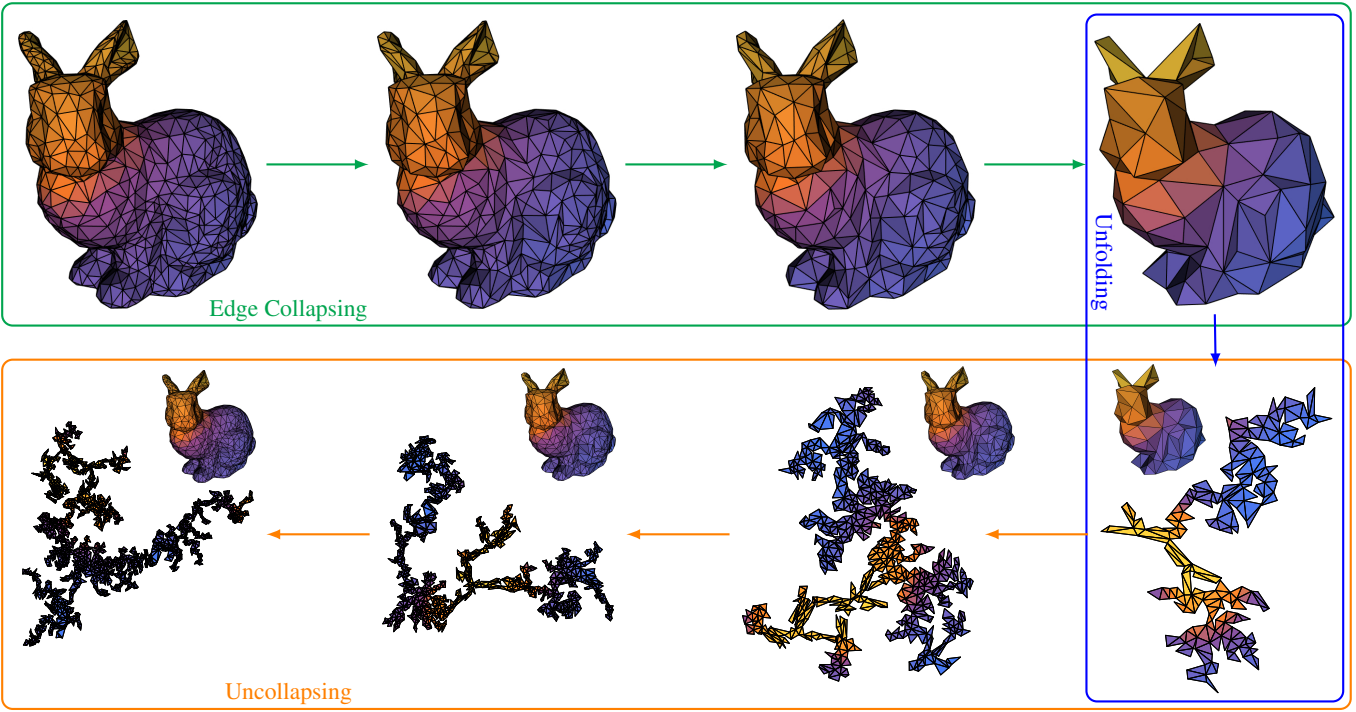}
  \caption{A visualization of our pipeline applied to the stanford bunny with 2000 faces.
           In the first phase, edges are collapsed (top row, green arrows).
           Then, the low resolution mesh is initially unfolded (right column, blue arrow).
           In the third phase, edges are uncollapsed while the unfolding is kept overlap-free (bottom row backwards, orange arrows).}
  \label{fig:Pipeline}
\end{figure*}

A common technique to overcome a difficult problem is to transform it, solve the transformed, and supposedly simpler, problem and then reverse-transform the result.
In our paper we ``transform'' an input mesh into a simplified version.
While there are many simplification methods~\cite[Chapter 7]{PolygonMeshProcessing}, for our application a method is needed which does small incremental steps, is robust and simple, and approximates the input well.
Edge collapsing~\cite{ProgressiveMeshes} fulfills all these criteria, which is why we chose this method for this paper.

\noindent
The pipeline of our paper can be split into three parts:

\begin{enumerate}
  \item Edge collapse a given mesh into a low resolution approximation.
  \item Unfold the approximation.
  \item Undo the collapsing, while keeping the unfolding overlap-free.
\end{enumerate}

This pipeline is illustrated in Figure~\ref{fig:Pipeline}.
A pseudo-code description of our algorithm is given in Algorithm~\ref{alg:PseudoCode} in Appendix~\ref{app:Pseudocode}.
We will call the third step of our pipeline \emph{uncollapsing}.
Our approach has multiple benefits compared to current techniques.
The lower the resolution of a mesh, the faster it can be unfolded.
If during the uncollapsing no significant new overlaps are introduced, the approach performs faster than a direct approach could.
On top of that, by allowing changes in the input, chances of overcoming not-unfoldability are increased.
Furthermore, any intermediate result of the progressive mesh refinement, through uncollapsing edges, results in an unfolded approximation.
For many applications, such an approximation is sufficient.

As an input, we restrict our meshes to be orientable, triangular manifold meshes.
The orientability and manifold constraints are necessary conditions for a mesh to be unfoldable.
In theory, it would be possible to allow arbitrary face-types.
For a mesh to be unfoldable, all faces must be flat, though.
When collapsing an edge in a non-triangular mesh, the remaining faces may not be flat anymore.
This problem could be overcome by adding an additional planarization step into the pipeline.
Since we want to investigate the effect of edge collapses in the topic of unfolding, we leave adding such a step for future work.
Thus, we restrict our input to be triangular.

\subsection{Edge Collapse Methods}
  \label{sec:CollapseMethods}

Edge collapse techniques are composed of an edge selection method and a vertex placement method.
The edge selection method is used to determine the order in which edges are collapsed, and the vertex placement rule determines the position of the vertex resulting from a collapse.
In our paper, we focus on quadric edge collapses~\cite{QuadricError}, which is a combination of a quadric edge selection and a quadric vertex placement method.
Quadric based mesh simplifications are known to approximate the input very well with few triangles.
Therefore, it is to be expected that during the uncollapsing part of our pipeline, branches in the unfolding will move very little.
To highlight this effect, we also compare to using a simple shortest edge - midpoint, as well as a shortest edge - quadric collapse strategy.
In contrast to quadrics, shortest edge - midpoint collapses are known to cause higher approximation errors, especially at sharp features of the mesh.
As shown in our evaluation (see Section~\ref{sec:Results}), this difference has a noticeable impact.

In the following, we will refer to the different edge collapse strategies by \emph{selection strategy short/placement strategy short}.
In particular, the quadric edge collapsing, with quadric vertex placement, will be named \emph{Q/Q}, the shortest edge - midpoint one \emph{SE/MP}, and the shortest edge - quadric one \emph{SE/Q}.

\subsection{Applying the Pipeline}
\label{sec:Pipeline}

In this section, we describe step by step how the pipeline of our algorithm works.

\subsubsection{Simplifying the Mesh}
\label{sec:Collapsing}

In the first stage of our algorithm, the edge collapse method is applied to simplify the input mesh.
Besides the edge collapse strategy, the number of faces to collapse to is an important value.
A too low resolution is not able to represent the same geometry as the input anymore, and will therefore lead to more computational overhead during the uncollapsing phase.
A too high resolution will result in more work during the unfolding step.
While in theory there is an optimum between these two extremes, in our experiments we found the effect on our pipeline to be marginal.
Hence we chose the targeted number of faces as
\begin{equation*}
  t_f = \frac{i_f}{10} + \sqrt{i_f} \cdot (1 + g),
\end{equation*}
with $i_f$ being the input number of faces and $g$ being the genus of the mesh.

\subsubsection{Unfolding}
\label{sec:Unfolding}

The initial unfolding is created using the Steepest Edge Unfolder~\cite{NetsOfPolyhedra}.
We resolve any remaining overlaps using the Tabu Unfolder (see Section~\ref{sec:BackgroundTabuUnfolding}).

\subsubsection{Reverse Transformation}
\label{sec:Uncollapsing}

\begin{figure}[ht]
  \begin{subfigure}{.49\linewidth}
    \includegraphics[width=\textwidth]{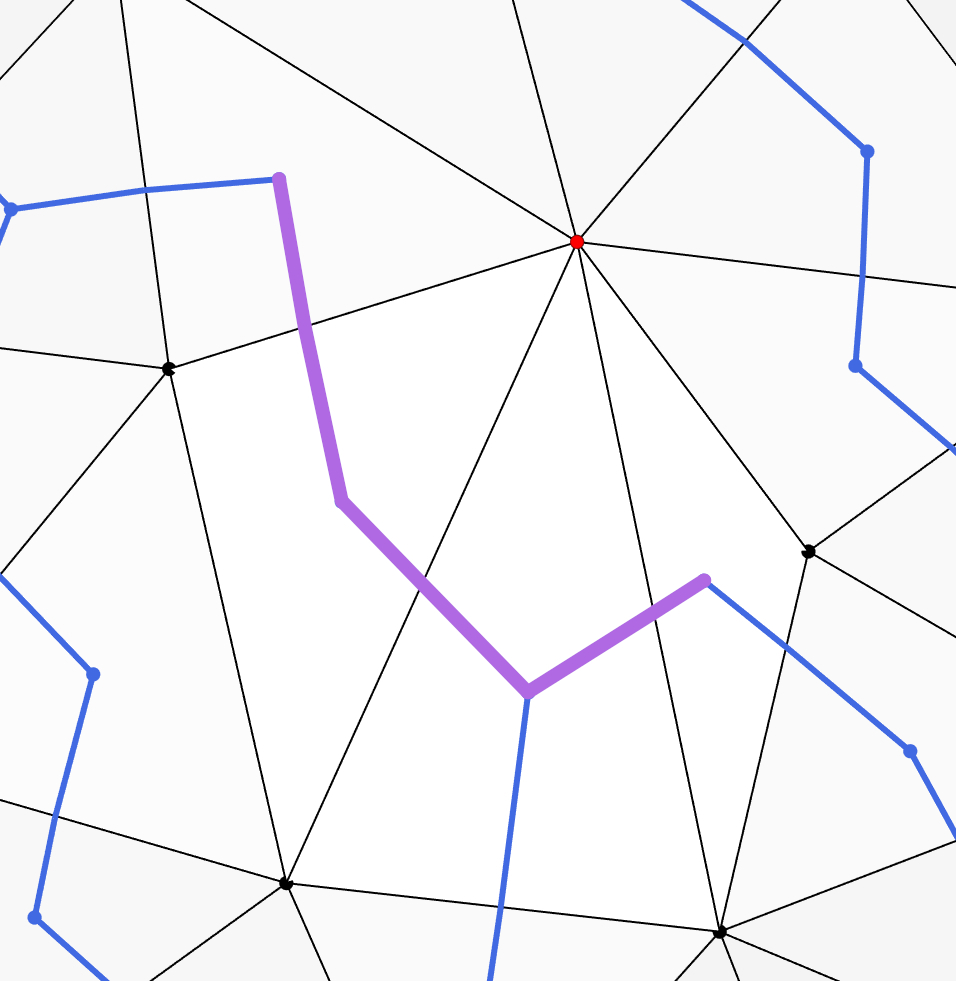}
    \caption{Before uncollapsing an edge.}
  \end{subfigure}
  \hfill
  \begin{subfigure}{.49\linewidth}
    \includegraphics[width=\textwidth]{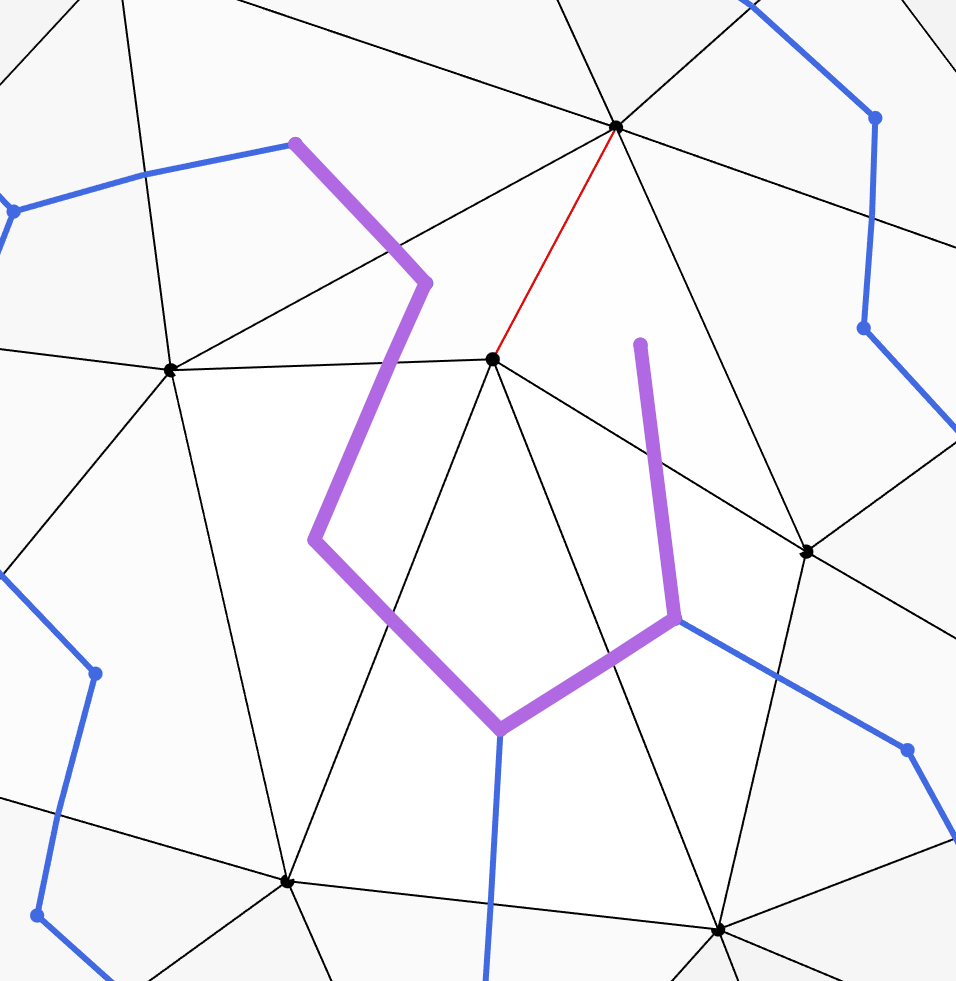}
    \caption{After uncollapsing an edge.}
    \label{fig:AfterUncollapseTree}
  \end{subfigure}
  \caption{Inserting triangles into the unfold-tree after uncollapsing an edge. The important parts of the unfold-tree are highlighted in purple and the uncollapsed edge is highlighted in red.}
  \label{fig:TriangleInsertion}
\end{figure}

In the last stage, the transformation (simplification) of the first stage is undone while keeping the unfolding overlap-free.
Uncollapsing an edge inserts one (in a boundary case) or two (in every other case) triangles into the mesh.
These triangles also need to be inserted into the current unfold-tree.
In particular, for each triangle (or node in the unfold-tree) two cases need to be distinguished:
Either the new node needs to be inserted inbetween two previously connected nodes or it can be inserted as a leaf.
In the latter case, we add the new node to a random neighbor in the unfold-tree.
The insertion of new faces generated by uncollapsing is illustrated in Figure~\ref{fig:TriangleInsertion}.

After inserting the new faces into the unfold-tree, our algorithm needs to re-evaluate the current configuration and does an overlap-check.
Since during insertion only the one-ring neighborhood of the split vertex has possibly changed, we can perform an overlap-test only on the subtrees determined by these faces~\cite[Section~4.7]{TabuUnfolding}, instead of the whole unfolding.
If overlaps are detected, we again use the Tabu Unfolder technique to resolve them, else we proceed to uncollapsing the next edge.
In case that during uncollapsing our algorithm can not find an overlap-free unfolding anymore, the last overlap-free unfolding is returned as an approximative result.

\section{Results}
\label{sec:Results}

\begin{figure}[ht]
  \centering
  \begin{subfigure}{.49\linewidth}
    \includegraphics[width=\textwidth]{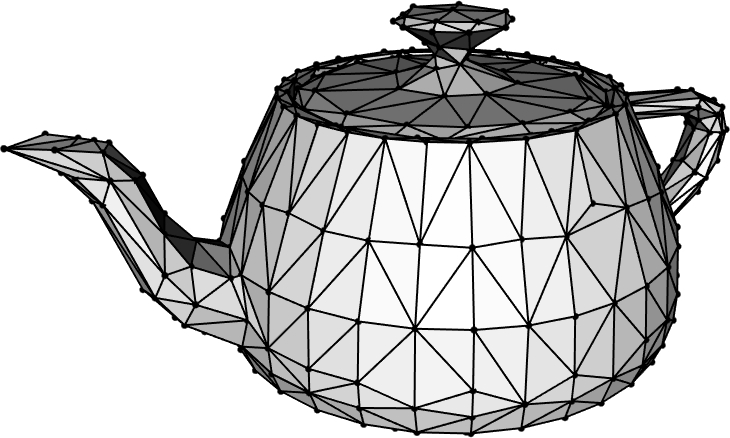}
    \caption{The 3D model.}
  \end{subfigure}
  \hfill
  \begin{subfigure}{.49\linewidth}
    \includegraphics[width=\textwidth]{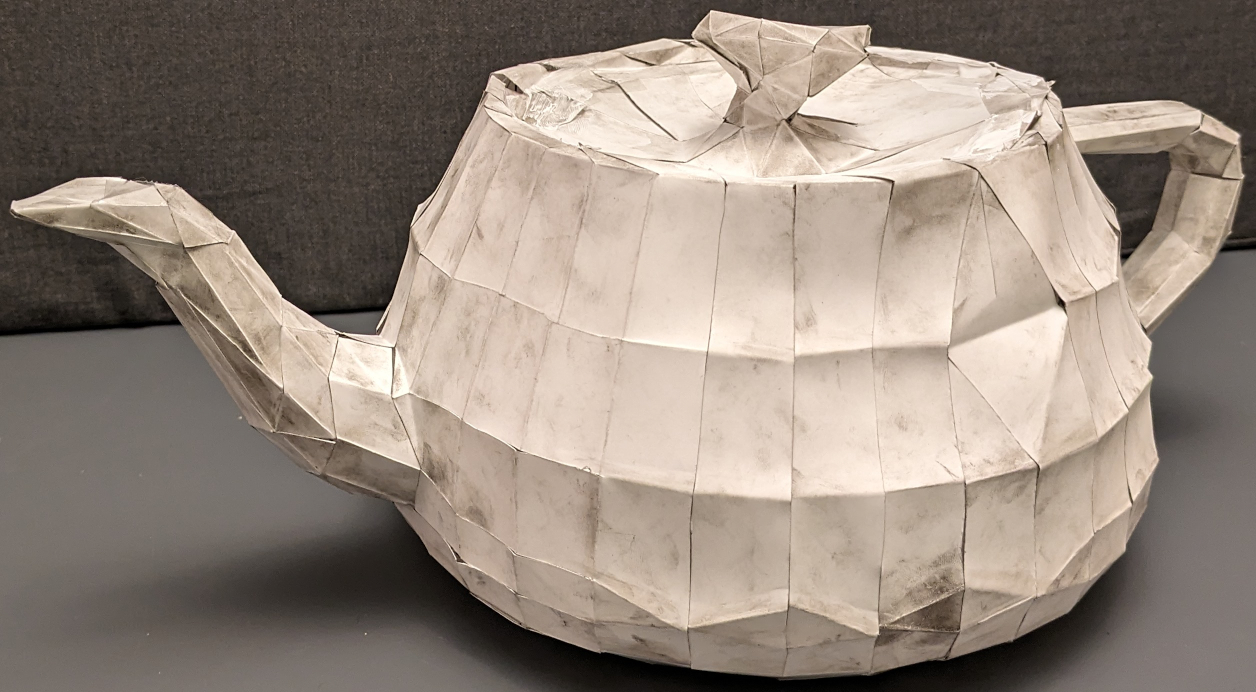}
    \caption{The folded model.}
  \end{subfigure}
  \caption{The Utah Tea Pot with 800 faces.}
  \label{fig:TeaPot}
\end{figure}

An example unfolding of the Utah Tea Pot and a refolded paper version are shown in Figure~\ref{fig:TeaPot}.

To determine the quality, efficiency, accuracy and reliability of our algorithm, we evaluated six different metrics:

\begin{itemize}
  \item \textbf{Coverages:}
  The coverage of an unfolding is defined as its summed triangle areas, divided by the area of its oriented bounding box.

  \item \textbf{Aspect ratios:}
  The aspect ratio of an unfolding is defined as the ratio of the oriented bounding box sides, such that the result is greater or equal to one.

  \item \textbf{Success rates:}
  A success is given, when a method is able to unfold the original model without approximation.
  In our pipeline, this is equivalent to finishing the reverse transformation completely.

  Separately, we also investigated the success rates including approximative results.
  
  \item \textbf{Approximation quality:}
  The Hausdorff distance between the original and the approximation, relative to the bounding box diagonal.

  \item \textbf{Timings:}
  Timings of successful unfoldings.
\end{itemize}

Coverages, and aspect ratios are very important metrics for the unfolding, since they determine the material waste when cutting, as well as usable paper formats.
Since we allow approximative results, the approximation quality is an important metric as well.
Timings and success rates do not affect the unfolding itself, but are obviously important metrics for the performance of the algorithm.
In our evaluation, we focused on the success rate including approximative results.
All metrics are discussed in the following subsections.
Eventually, we also discuss any limitations.

For our evaluation, we chose a subset of the Thingi10k dataset~\cite{Thingi10k}.
To match our constraints, we only kept meshes which were manifold, consisted of a single component and were representable with 100 faces and were made of at least 2000 faces.
Our subset consisted of 2,800 meshes, which we represented in ten different resolutions (using 100, 200, 400, 600, 800, 1000, 1250, 1500, 1750, and 2000 faces).
We evaluated our algorithm in three variants, the Q/Q, SE/MP, and SE/Q one as described in Section~\ref{sec:CollapseMethods}.
As a comparison basis, we chose the Tabu Unfolder.

\subsection{Coverage, Aspect Ratios, and Success Rates}
\label{sec:AspectRatiosAndCoverage}

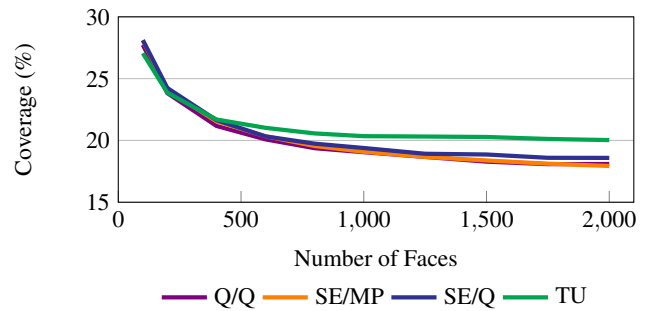
\begin{figure}[ht]
  \centering
  \begin{center}
\begin{tikzpicture}
\begin{axis}[
    xmin = 0, xmax = 2100,
    ymin = 15, ymax = 30,
    ymajorgrids,
    ytick style={draw=none},
    xtick pos=bottom,
    major grid style = {lightgray},
    minor grid style = {lightgray!25},
    width = \linewidth,
    height = .48\linewidth,
    xlabel = Number of Faces,
    ylabel = Coverage (\%),
    legend style={at={(0.5, -0.4)}, anchor = north, draw=none, legend columns = 4}
]

\addplot[violet, ultra thick] table [x ={x}, y = {Coverage}] {data/Performance-Dataset/Q_Q.dat};
\addlegendentry{Q/Q}

\addplot[orange, ultra thick] table [x = {x}, y = {Coverage}] {data/Performance-Dataset/SE_MP.dat};
\addlegendentry{SE/MP}

\addplot[Blue, ultra thick] table [x = {x}, y = {Coverage}] {data/Performance-Dataset/SE_Q.dat};
\addlegendentry{SE/Q}

\addplot[Green, ultra thick] table [x = {x}, y = {Coverage}] {data/Performance-Dataset/TU.dat};
\addlegendentry{TU}

\end{axis}
\end{tikzpicture}
\end{center}
  \caption{The mean coverages of our algorithm compared to Tabu Unfolding.}
  \label{fig:Coverage}
\end{figure}

The coverages, shown in Figure~\ref{fig:Coverage}, of our pipeline and the Tabu Unfolder are almost the same, especially the general trend is similar.
Since our method, as well as the Tabu Unfolder, aim for single-patched unfoldings, coverages are naturally lower compared to segmenting approaches.

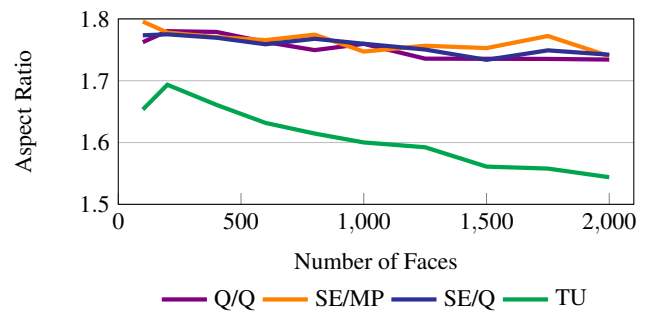
\begin{figure}[ht]
  \centering
  \begin{center}
\begin{tikzpicture}
\begin{axis}[
    xmin = 0, xmax = 2100,
    ymin = 1.5, ymax = 1.8,
    ymajorgrids,
    ytick style={draw=none},
    xtick pos=bottom,
    major grid style = {lightgray},
    minor grid style = {lightgray!25},
    width = \linewidth,
    height = .48\linewidth,
    xlabel = Number of Faces,
    ylabel = Aspect Ratio,
    legend style={at={(0.5, -0.4)}, anchor = north, draw=none, legend columns = 4}
]

\addplot[violet, ultra thick] table [x ={x}, y = {AspectRatio}] {data/Performance-Dataset/Q_Q.dat};
\addlegendentry{Q/Q}

\addplot[orange, ultra thick] table [x = {x}, y = {AspectRatio}] {data/Performance-Dataset/SE_MP.dat};
\addlegendentry{SE/MP}

\addplot[Blue, ultra thick] table [x = {x}, y = {AspectRatio}] {data/Performance-Dataset/SE_Q.dat};
\addlegendentry{SE/Q}

\addplot[Green, ultra thick] table [x = {x}, y = {AspectRatio}] {data/Performance-Dataset/TU.dat};
\addlegendentry{TU}

\end{axis}
\end{tikzpicture}
\end{center}
  \caption{The mean aspect ratios of our algorithm compared to Tabu Unfolding.}
  \label{fig:AspectRatios}
\end{figure}

In contrast, the aspect ratios, as shown in Figure~\ref{fig:AspectRatios}, differ both in their values as well as their overall trend.
While the Tabu Unfolder shows a clear downward trend, our method results in an almost constant aspect ratio close to 1.8.
Thus our progressive mesh unfolding is easier to use in a fixed production pipeline, since the same paper type can be used for any resolution.

\begin{figure}[ht]
  \centering
  \begin{center}
\begin{tikzpicture}
\begin{axis}[
    xmin = 0, xmax = 2100,
    ymin = 98.5, ymax = 100,
    ymajorgrids,
    ytick style={draw=none},
    xtick pos=bottom,
    major grid style = {lightgray},
    minor grid style = {lightgray!25},
    axis x line*=bottom,
    width = \linewidth,
    height = .48\linewidth,
    xlabel = Number of Faces,
    ylabel = Success Rate Approx (\%),
    legend style={at={(0.5, -0.4)}, anchor = north, draw=none, legend columns = 4}
]

\addplot[violet, ultra thick] table [x ={x}, y = {SuccessRateApprox}] {data/Performance-Dataset/Q_Q.dat};
\addlegendentry{Q/Q}

\addplot[orange, ultra thick] table [x = {x}, y = {SuccessRateApprox}] {data/Performance-Dataset/SE_MP.dat};
\addlegendentry{SE/MP}

\addplot[Blue, ultra thick] table [x = {x}, y = {SuccessRateApprox}] {data/Performance-Dataset/SE_Q.dat};
\addlegendentry{SE/Q}

\addplot[Green, ultra thick] table [x = {x}, y = {SuccessRateApprox}] {data/Performance-Dataset/TU.dat};
\addlegendentry{TU}

\end{axis}
\end{tikzpicture}
\end{center}
  \caption{The success rates of our algorithm compared to Tabu Unfolding including approximative results.}
  \label{fig:SuccessRatesApprox}
\end{figure}
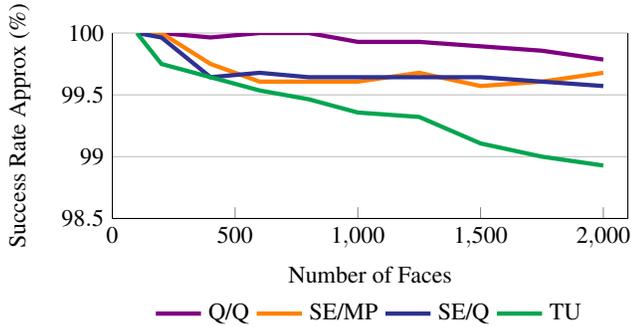

Finally, within the success rates including approximative results, a clear difference is visible as shown in Figure~\ref{fig:SuccessRatesApprox}.
The Tabu Unfolder shows an almost linear downward trend.
Our method in the SE variant with both placement methods first decay and then show a constant trend.
The Q/Q variant of our algorithm shows the best reliability throughout.
Success rates without approximation are shown in Figure~\ref{fig:SuccessRates} in Appendix~\ref{app:SuccessRates}.
Thus we have a clear benefit in the success rates using our progressive mesh unfolding approach.

\subsection{Approximative Results}
\label{sec:ApproximativeResults}

\begin{figure}[ht]
  \centering
  \begin{center}
\begin{tikzpicture}
\begin{axis}[
    xmin = 0, xmax = 2100,
    ymin = 0, ymax = 9,
    ymajorgrids,
    ytick style={draw=none},
    xtick pos=bottom,
    major grid style = {lightgray},
    minor grid style = {lightgray!25},
    width = \linewidth,
    height = .48\linewidth,
    xlabel = Number of Faces,
    ylabel = Hausdorff Distance (\%),
    legend style={at={(0.5, -0.35)}, anchor = north, draw=none, legend columns = 3}
]

\addplot[violet, ultra thick] table [x ={x}, y = {Hausdorffschnitt}] {data/Hausdorff/Q_Q.dat};
\addlegendentry{Q/Q}

\addplot[orange, ultra thick] table [x = {x}, y = {Hausdorffschnitt}] {data/Hausdorff/SE_MP.dat};
\addlegendentry{SE/MP}

\addplot[Blue, ultra thick] table [x = {x}, y = {Hausdorffschnitt}] {data/Hausdorff/SE_Q.dat};
\addlegendentry{SE/Q}

\end{axis}
\end{tikzpicture}
\end{center}
  \caption{Mean relative Hausdorff distances over all shared approximative results for each variant of our algorithm.}
  \label{fig:HausdorffSchnitt}
\end{figure}
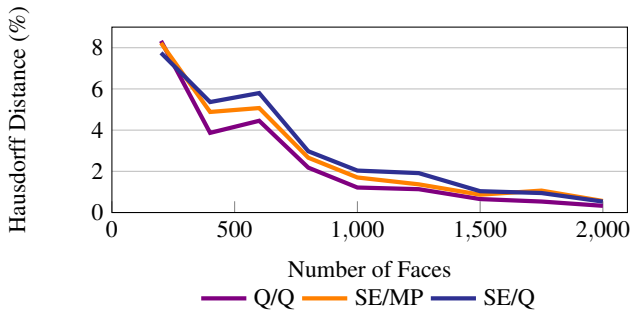

We investigated the approximation quality of our algorithm by measuring the Hausdorff distance between approximative results and their originals, relative to the respective bounding box diagonal.
To make these numbers comparable, we first evaluated the approximation quality only for those meshes, which all three variants approximated.
These results are shown in Figure~\ref{fig:HausdorffSchnitt}.
As expected, the Q/Q variant of our algorithm shows the best approximation quality.
Notably, all approximation errors reach almost zero percent at 2,000 faces.
This, in combination with the high success rates (see Figure~\ref{fig:SuccessRatesApprox}) make our algorithm a good choice for unfolding meshes with a few thousand faces.

\begin{figure}[ht]
  \centering
  \includegraphics[width=\linewidth]{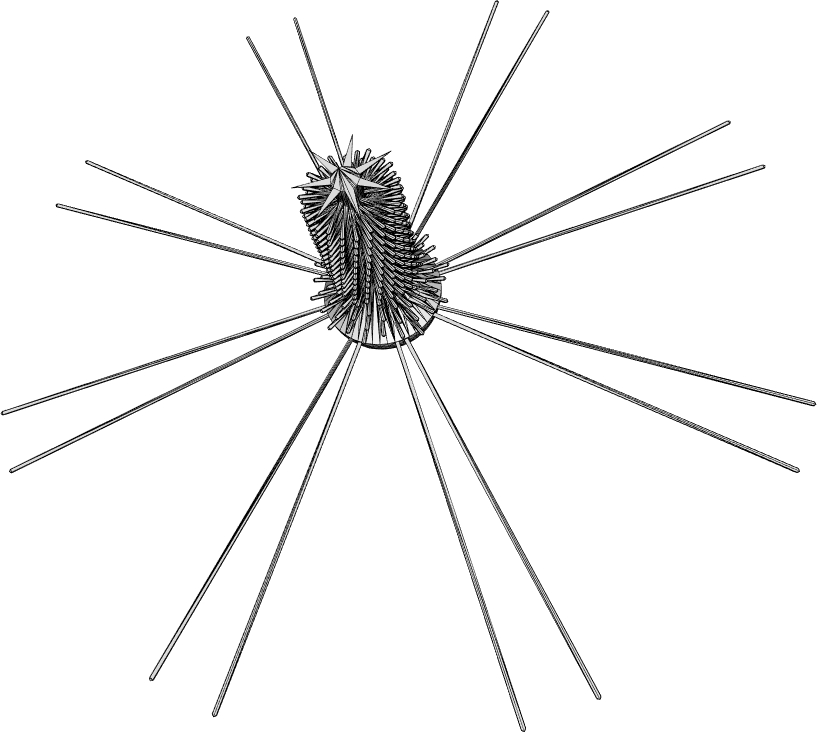}
  \caption{An example mesh (Thingi-ID: 466802) with complex geometry. Here, our approach }
  \label{fig:NastyExample}
\end{figure}

As shown in Figure~\ref{fig:SuccessRatesApprox}, our algorithm succeeds more often in the Q/Q variant, than in the SE/MP one.
The cases where this happens have a complex geometry in common, which can not be well approximated with few faces.
Such an example is shown in Figure~\ref{fig:NastyExample}.
For such meshes, the SE/MP variant either fails to produce any result, or produces a very high approximation error.
In contrast, the Q/Q variant succeeds more often on those meshes, but sometimes also produces a high approximation error.
E.g., in the mesh shown in Figure~\ref{fig:NastyExample}, if some of the long spikes are lost, the error will be very high, regardless of how well the remaining mesh is approximated.

\subsection{Timings}
\label{sec:Performance}

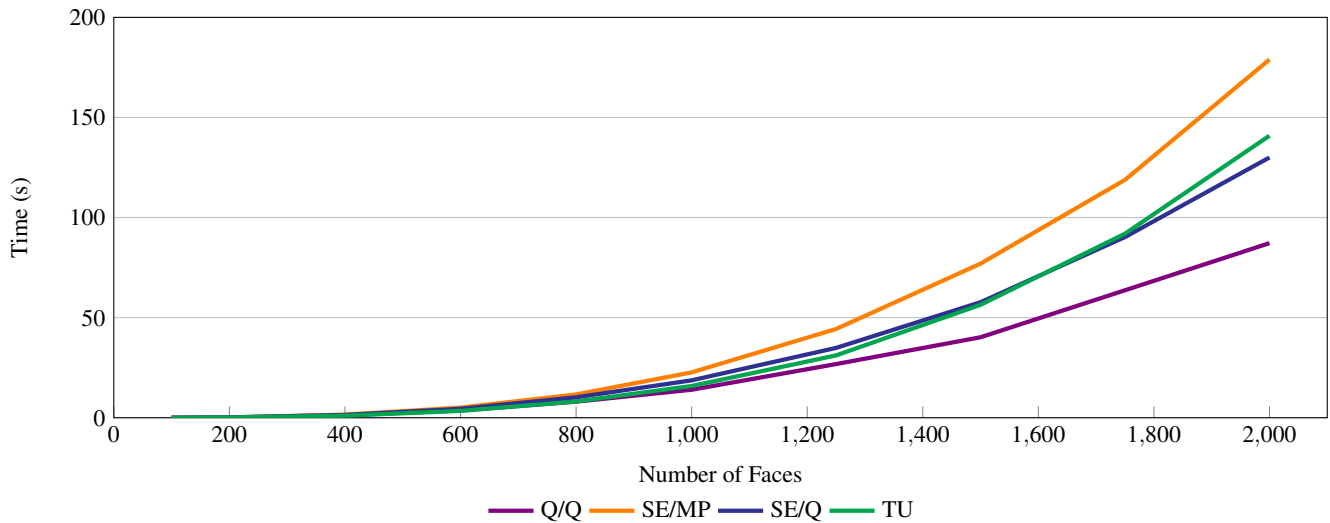
\begin{figure*}[ht]
  \begin{center}
\begin{tikzpicture}
\begin{axis}[
    xmin = 0, xmax = 2100,
    ymin = 0, ymax = 200,
    ymajorgrids,
    ytick style={draw=none},
    xtick pos=bottom,
    major grid style = {lightgray},
    minor grid style = {lightgray!25},
    width = \linewidth,
    height = .39\linewidth,
    xlabel = Number of Faces,
    ylabel = Time (s),
    legend style={at={(axis description cs:0.3,-0.18)}, anchor=north west, draw=none, legend columns = 4}
]

\addplot[violet, ultra thick] table [x ={x}, y = {MedianTime}] {data/Performance-Dataset/Q_Q.dat};
\addlegendentry{Q/Q}

\addplot[orange, ultra thick] table [x = {x}, y = {MedianTime}] {data/Performance-Dataset/SE_MP.dat};
\addlegendentry{SE/MP}

\addplot[Blue, ultra thick] table [x = {x}, y = {MedianTime}] {data/Performance-Dataset/SE_Q.dat};
\addlegendentry{SE/Q}

\addplot[Green, ultra thick] table [x = {x}, y = {MedianTime}] {data/Performance-Dataset/TU.dat};
\addlegendentry{TU}

\end{axis}
\end{tikzpicture}
\end{center}
  \caption{Median unfold timings of our method compared to Tabu Unfolding.}
  \label{fig:PerformanceDataset}
\end{figure*}

\begin{figure}[ht]
  \centering
  \begin{subfigure}{\linewidth}
    \includegraphics[width=\textwidth]{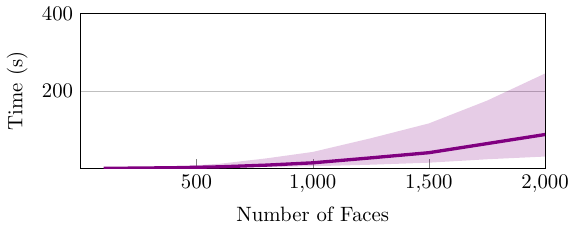}
    \caption{Q/Q}
  \end{subfigure}
  \begin{subfigure}{\linewidth}
    \includegraphics[width=\textwidth]{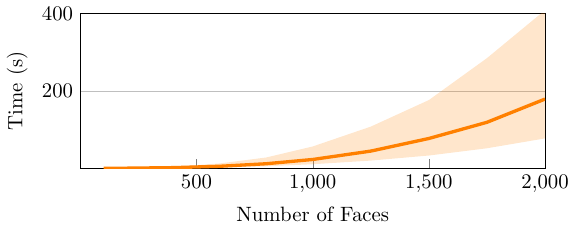}
    \caption{SE/MP}
  \end{subfigure}
  \begin{subfigure}{\linewidth}
    \includegraphics[width=\textwidth]{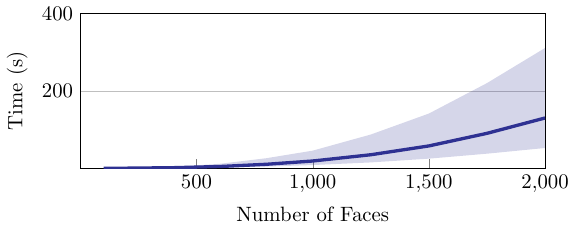}
    \caption{SE/Q}
  \end{subfigure}
  \begin{subfigure}{\linewidth}
    \includegraphics[width=\textwidth]{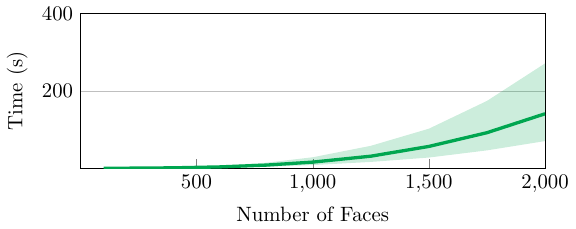}
    \caption{TU}
  \end{subfigure}
  \caption{A side-by-side comparison of the timings for our algorithm in different variants, and Tabu Unfolding, including distribution-indicators. The line for each graph is the median time, the area around the line marks the value at 25\% and 75\% of the recorded timings.}
  \label{fig:TimeDistributions}
\end{figure}

All timings have been measured on a machine with an i7-10700K CPU (3.8GHz) and 128GB RAM running Linux.
All implementations were written in C++ and were executed without parallelization.
We executed our algorithm (in its three variants), as well as Tabu Unfolding, on each mesh in the test set in each resolution once.
Even though the test set is large, each model has only been unfolded once and thus average times are still affected by outliers.
Therefore, we analyze the median timings (see Figure~\ref{fig:PerformanceDataset}), which are much less sensitive to outliers.
In Figure~\ref{fig:TimeDistributions} a side-by-side comparison of the timings including distribution-indicators are shown.
Detailed results are shown in Tables~\ref{tab:DetailedTimings} in Appendix~\ref{app:Results}.

We can observe that our algorithm performs faster than the Tabu Unfolder.
Using a quadric edge collapse strategy is the key factor to achieve faster timings, even though the Tabu Unfolder is used as the underlying overlap removal method.
As mentioned in Section~\ref{sec:CollapseMethods}, quadrics are known to produce very good approximations.
These low resolution approximations are faster to unfold, since they contain less faces.
If the geometry does (almost) not change, when uncollapsing an edge, the unfolding will also change very little.
This effect can be seen in Figure~\ref{fig:Pipeline}.
Besides a global rotation of the branches, the unfolding mostly changes by adding more detail and very little in the general shape of its branches.

\subsection{Limitations}
\label{sec:Limitations}

Some well-known examples, which are not-unfoldable~\cite[Chapter~22.4]{GeometricFoldingAlgorithms}, are made from very few triangles and become unfoldable in higher resolutions.
Since our algorithm aims to unfold a low resolution mesh, which approximates the same geometry as the high resolution shape, it may happen that this low resolution mesh is not-unfoldable, even though its high resolution counterpart would be.
In such a case, our algorithm fails to produce a result.
The results shown in this paper suggest this case occurs very rarely, and that our strategy is generally viable.

Additionally, there is no guarantee that an approximative result is not-unfoldable in its original resolution.
One step in the uncollapsing being not-unfoldable does not necessarily imply that the original is not-unfoldable.

\section{Conclusion and Future Work}
\label{sec:Conclusion}

In this work, we presented an algorithm, which unfolds a given triangular mesh of arbitrary genus, by first collapsing it into a low resolution approximation, unfolds this approximation, and then uncollapses the approximation back to the original input, while keeping the unfolding overlap-free.
We showed that this approach performs faster than other current approaches, while also providing higher reliability.
Additionally, our approach is also able to handle not-unfoldable input by approximation.
In contrast to other works, our progressive mesh unfolding does not need to segment the result.
As argued in Section~\ref{sec:Results}, a shape-preserving placement strategy like the quadric one is essential for the advantages our algorithm provides.

Currently, we only investigated unfolding of triangle meshes.
In the future, we would like to investigate ways to apply our algorithm to other polygonal meshes.

  \bibliographystyle{arxiv}
  \bibliography{bibliography}

\newpage
\appendix

\section{Pseudo-Code}
\label{app:Pseudocode}
  \begin{algorithm}[ht]
    \begin{algorithmic}
      \Require $C$ \Comment{$C$ is the decimator, applying a collapse strategy}
      \Require $S$ \Comment{$S$ is the 2D overlap solver, i.e. the Tabu Unfolder}
      \Require $M$ \Comment{$M$ is the mesh to unfold}
      \Function{CollapseUnfolding}{$C, S, M$}
        \State $tnf \gets \frac{M.numF()}{10} + \sqrt{M.numF()} \cdot M.genus()$
        \State $M \gets C.collapse(tnf)$ \Comment{Section~\ref{sec:Collapsing}}
        \State $U \gets createInitialUnfolding(M)$ \Comment{Section~\ref{sec:Unfolding}}
        \If{$!S.resolveOverlaps(U)$} \Comment{\cite{TabuUnfolding}}
          \State \Return $-1$ \Comment{Failed case}
        \EndIf
        \While{$!C.finishedUncollapsing()$} \Comment{Section~\ref{sec:Uncollapsing}}
          \State $U' \gets U$
          \State $U.setVertices(C.uncollapseEdge())$
          \If{$!S.resolveOverlaps(U)$} \Comment{\cite{TabuUnfolding}}
            \State $U \gets U'$ \Comment{Restore last valid state}
            \State \Return $1$ \Comment{Approximative result}
          \EndIf
        \EndWhile
        \State \Return $0$ \Comment{Success}
      \EndFunction
    \end{algorithmic}
    \caption{Progressive Mesh Unfolding}
    \label{alg:PseudoCode}
  \end{algorithm}

\section{Success Rates}
\label{app:SuccessRates}

\begin{figure}[ht]
  \centering
  \begin{center}
\begin{tikzpicture}
\begin{axis}[
    xmin = 0, xmax = 2100,
    ymin = 96, ymax = 100,
    ymajorgrids,
    ytick style={draw=none},
    xtick pos=bottom,
    major grid style = {lightgray},
    minor grid style = {lightgray!25},
    axis x line*=bottom,
    width = \linewidth,
    height = .48\linewidth,
    xlabel = Number of Faces,
    ylabel = Success Rate (\%),
    legend style={at={(0.5, -0.4)}, anchor = north, draw=none, legend columns = 4}
]

\addplot[violet, ultra thick] table [x ={x}, y = {SuccessRate}] {data/Performance-Dataset/Q_Q.dat};
\addlegendentry{Q/Q}

\addplot[orange, ultra thick] table [x = {x}, y = {SuccessRate}] {data/Performance-Dataset/SE_MP.dat};
\addlegendentry{SE/MP}

\addplot[Blue, ultra thick] table [x = {x}, y = {SuccessRate}] {data/Performance-Dataset/SE_Q.dat};
\addlegendentry{SE/Q}

\addplot[Green, ultra thick] table [x = {x}, y = {SuccessRate}] {data/Performance-Dataset/TU.dat};
\addlegendentry{TU}

\end{axis}
\end{tikzpicture}
\end{center}
  \caption{The success rates of our algorithm compared to Tabu Unfolding excluding approximative results.}
  \label{fig:SuccessRates}
\end{figure}
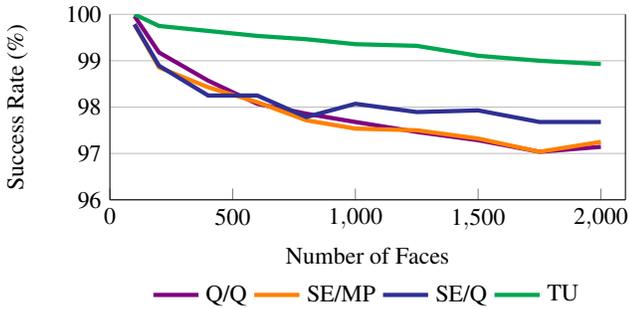

\section{Detailed Results}
\label{app:Results}

{
\begin{table}[!hb]
  \centering
  \footnotesize
  \begin{tabular}{lr|rrrr}
    Value                               & |F|  &        Q/Q &     SE/MP  &       SE/Q &         TU  \\
    \hline
                                        &      & \multicolumn{4}{c}{\textbf{Timings (s)}}           \\
    \hline
    Min                                 &  all &      0.000 &      0.000 &      0.000 &      0.000  \\
    \hline
    \multirow{10}{*}{Median}            &  100 &      0.018 &      0.020 &      0.018 &      0.015  \\
                                        &  200 &      0.166 &      0.174 &      0.168 &      0.135  \\
                                        &  400 &      1.300 &      1.544 &      1.408 &      1.020  \\
                                        &  600 &      3.623 &      5.058 &      4.464 &      3.432  \\
                                        &  800 &      8.037 &     11.630 &     10.205 &      8.129  \\
                                        & 1000 &     13.978 &     22.656 &     18.689 &     15.857  \\
                                        & 1250 &     26.812 &     44.340 &     34.885 &     31.149  \\
                                        & 1500 &     40.205 &     76.985 &     57.719 &     56.416  \\
                                        & 1750 &     63.685 &    118.837 &     90.264 &     91.944  \\
                                        & 2000 &     87.213 &    178.930 &    130.008 &    140.888  \\
    \hline
    \multirow{10}{*}{Mean}              &  100 &      0.049 &      0.050 &      0.049 &      0.030  \\
                                        &  200 &      0.544 &      0.598 &      0.575 &      0.282  \\
                                        &  400 &      6.170 &      6.031 &      5.724 &      2.257  \\
                                        &  600 &     22.027 &     21.116 &     21.636 &      6.940  \\
                                        &  800 &     52.701 &     50.592 &     45.994 &     19.106  \\
                                        & 1000 &     90.594 &     98.783 &     87.323 &     42.305  \\
                                        & 1250 &    168.444 &    199.345 &    160.896 &     88.723  \\
                                        & 1500 &    248.333 &    324.464 &    285.741 &    141.768  \\
                                        & 1750 &    433.656 &    549.684 &    457.902 &    246.291  \\
                                        & 2000 &    659.311 &    728.359 &    626.908 &    352.958  \\
    \hline
    \multirow{10}{*}{Max}               &  100 &      2.396 &      2.082 &      2.983 &      1.104  \\
                                        &  200 &     23.006 &     25.837 &     27.673 &     23.238  \\
                                        &  400 &    248.122 &    253.816 &    247.668 &    198.557  \\
                                        &  600 &    892.522 &    800.405 &    903.418 &    607.050  \\
                                        &  800 &   2090.394 &   2249.449 &   2228.795 &   1150.672  \\
                                        & 1000 &   4583.610 &   4363.279 &   4517.269 &   3586.101  \\
                                        & 1250 &   7350.622 &   9175.251 &   8201.845 &   7747.810  \\
                                        & 1500 &  13703.176 &  15587.310 &  16143.076 &   8964.870  \\
                                        & 1750 &  24035.986 &  26133.299 &  26603.642 &  11152.966  \\
                                        & 2000 &  37216.614 &  39559.031 &  39313.610 &  20322.664  \\
    \hline
                                        &      & \multicolumn{4}{c}{\textbf{Success Rates (\%)}}    \\
    \hline
    \multirow{10}{*}{Regular}           &  100 &     99.964 &     99.786 &     99.786 &    100.000  \\
                                        &  200 &     99.179 &     98.857 &     98.893 &     99.750  \\
                                        &  400 &     98.571 &     98.429 &     98.250 &     99.643  \\
                                        &  600 &     98.071 &     98.107 &     98.250 &     99.536  \\
                                        &  800 &     97.857 &     97.714 &     97.786 &     99.464  \\
                                        & 1000 &     97.679 &     97.536 &     98.071 &     99.357  \\
                                        & 1250 &     97.464 &     97.500 &     97.893 &     99.321  \\
                                        & 1500 &     97.286 &     97.321 &     97.929 &     99.107  \\
                                        & 1750 &     97.036 &     97.036 &     97.679 &     99.000  \\
                                        & 2000 &     97.143 &     97.250 &     97.679 &     98.929  \\
    \hline
    \multirow{10}{*}{Approx}            &  100 &    100.000 &    100.000 &    100.000 &    100.000  \\
                                        &  200 &    100.000 &    100.000 &     99.964 &     99.750  \\
                                        &  400 &     99.964 &     99.750 &     99.643 &     99.643  \\
                                        &  600 &    100.000 &     99.607 &     99.679 &     99.536  \\
                                        &  800 &    100.000 &     99.607 &     99.643 &     99.464  \\
                                        & 1000 &     99.929 &     99.607 &     99.643 &     99.357  \\
                                        & 1250 &     99.929 &     99.679 &     99.643 &     99.321  \\
                                        & 1500 &     99.893 &     99.571 &     99.643 &     99.107  \\
                                        & 1750 &     99.857 &     99.607 &     99.607 &     99.000  \\
                                        & 2000 &     99.786 &     99.679 &     99.571 &     98.929  \\
  \end{tabular}
  \caption{\normalsize Detailed timings and success rates for all variants of our algorithm, as well as Tabu Unfolding.}
  \label{tab:DetailedTimings}
\end{table}
}

\end{document}